# Artificial Human Lecturers: Initial Findings From Asia's First AI Lecturers in Class to Promote Innovation in Education

CHING CHRISTIE PANG, Hong Kong University of Science and Technology, China
YAWEI ZHAO, Hong Kong University of Science and Technology (Guangzhou), China
ZHIZHUO YIN, Hong Kong University of Science and Technology (Guangzhou), China
JIA SUN, Hong Kong University of Science and Technology (Guangzhou), China
REZA HADI MOGAVI, University of Waterloo, Canada
PAN HUI, Hong Kong University of Science and Technology (Guangzhou), China

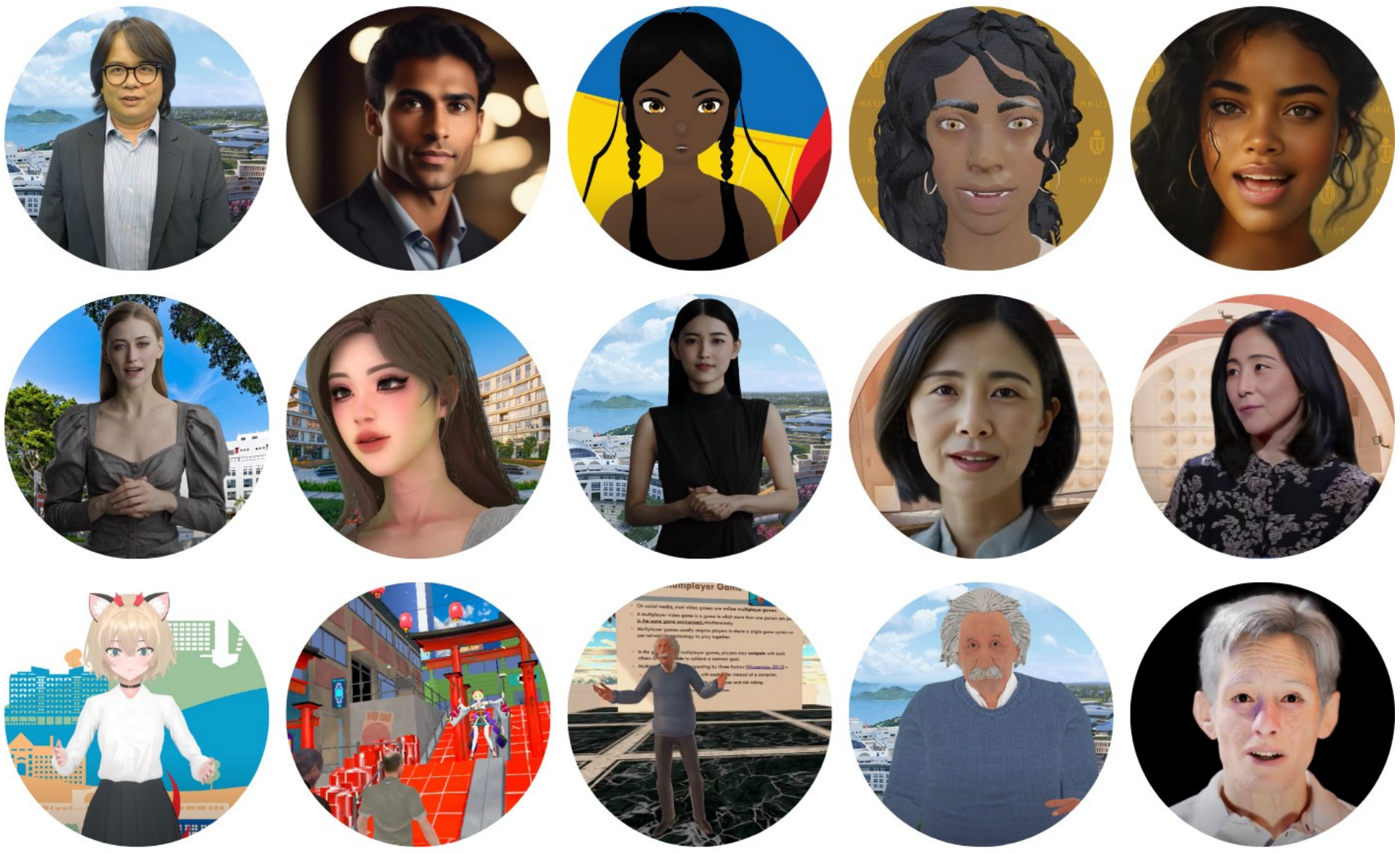

Fig. 1. The introduction of AI digital teachers has revolutionized the way students learn. Using generative AI technology, we developed ten pairs of "AI Lecturers" with diverse personas and backgrounds.

In recent years, artificial intelligence (AI) has become increasingly integrated into education, reshaping traditional learning environments. Despite this, there has been limited investigation into fully operational artificial human lecturers. To the best of our knowledge, our paper presents the world's first study examining their deployment in a real-world educational setting. Specifically, we investigate the use of "digital teachers," AI-powered virtual lecturers, in a postgraduate course at the Hong Kong University of Science and Technology (HKUST). Our study explores how features such as *appearance*, *non-verbal cues*, *voice*, and *verbal expression* impact students' learning experiences. Findings suggest that students highly value *naturalness*, *authenticity*, and *interactivity* in digital teachers, highlighting areas for improvement, such as *increased responsiveness*, *personalized avatars*, and *integration with larger learning platforms*. We conclude that digital teachers have significant potential to enhance education by providing a more *flexible*, *engaging*, *personalized*, and *accessible* learning experience for students.

## 1 INTRODUCTION

The global teacher shortage is a pressing issue supported by several alarming trends. In February 2024, UNESCO predicted that an extra 44 million teachers would be needed globally by 2030 to attain universal education, with 17 million teachers being needed in Sub-Saharan Africa alone [4]. Economic disparities exacerbate this crisis. According to the Global Partnership for Education [1], many low-income nations devote less than 3% of their GDP to education. This leads to inadequate salaries, which push teachers — especially those in countries like Zimbabwe — to look for better opportunities abroad. Demographic changes further complicate the situation. The World Bank anticipates a 30% increase in student enrollment in the Middle East and North Africa by 2030 [66], necessitating a significant rise in qualified educators. Additionally, retention challenges plague the profession globally. According to research published in Educational Researcher [43], nearly 50% of new teachers leave within the first five years due to factors such as high workloads and a lack of support. This is further corroborated by the National Center for Education Statistics [2] which found that 8% of teachers in the US leave their jobs each year. Over 32,000 educators left Korea before they reached retirement age [37], while 43,500 teachers left the UK state-funded education system in 2022–2023 [3].

These examples illuminate the acute global teacher shortage, prompting our foundational motivated question in this work: **How can modern technological advances help alleviate the short-staffed education sector?** To answer this question, we conduct exploratoty field study by introducing 10 pairs of AI-powered digital lecturers into a graduate course that lasts for a semester. This study is based on observation and data analysis that was obtained from quantitative sequential surveys with 26 students and qualitative data from the in-depth interviews with 15 of those students.

This work on digital teachers is important for at least three reasons. First, it provides practitioners with methods that contribute to the development of effective pedagogical strategies, enabling educators to integrate digital teachers into their teaching practices. Second, it offers an artifact contribution by presenting a comprehensive AIGC (Artificial Intelligence Generated Content) stratagem specifically designed for digital teachers, thereby enhancing the educational landscape. Third, it introduces Asia's first "AI Lecturers" in classrooms, promoting teaching innovation. More than 150 agencies have reported on the innovative strategy, receiving extensive media coverage. This initiative has the potential to inspire further relevant research and practices that address the existing teacher shortage problem.

This study is significant and difficult for a number of reasons. First, practical obstacles make digital teaching approaches in field research difficult. These include a scarcity of VR headsets and platforms, among other essential equipment, and a small pool of willing student participants. Second, there are a lot of technological hurdles in the creation of completely effective AI-powered digital teachers. These include the need to create lifelike avatars, generate intelligible speech, accurately sync lips, and incorporate a variety of styles and gestures. Third, since individual variances in perception can impact user acceptability and engagement, the subjectivity inherent in user preferences and aesthetic sensibility further complicates the study. To effectively address the worldwide teacher shortage and develop digital teaching solutions, it is imperative to tackle these difficulties.

In order to gather a general sense of students' opinions and views of digital instructors, we employed the Technology Acceptance Model (TAM) [19] in our survey. We eventually settled on four distinct criteria for digital teachers that influence student perceptions: (i) **Appearance and Character Choice**, (ii) **Non-verbal Cues**, (iii) **Voice and Verbal Expression**, and (iv) **Variety and Novelty**. These parameters were chosen because they encompass all of the crucial components that affect user acceptability and engagement. Appearance and Character Choice affect initial impressions, while Non-verbal Cues enhance communication and relatability. Voice and Verbal Expression contribute to the clarity and emotional resonance of interactions, and Variety and Novelty ensure that the digital teaching experience remains engaging and dynamic. When taken as a whole, these standards offer a thorough foundation for comprehending how students view and communicate with digital instructors. Therefore, we focus our analysis on two research questions:

- **RQ1:** How does the variation of the four parameters in digital teachers impact students' experiences?
- **RQ2:** According to students, what are the potential areas for improvement in the implementation of digital teachers, and how do they view the future of digital teachers in education?

The result is our design guidelines and perceivable future improvement for the use of digital teachers. Our findings show that differences in the four parameters have a major influence on students' dimensional experiences. Regarding the employment of digital teachers in the classroom, most participants were "optimistic", emphasizing the innovative, effective, productive, and visionary potential of new technologies. Nonetheless, common opinions also expressed curiosity, indifference, and concern over the use of AI lecturers. Students pointed out several possible areas where the use of digital instructors may be improved, including the requirement for more individualized interactions and better flexibility to accommodate different learning styles. Participants in a discussion on the potential of digital instructors in the classroom expressed hope for the technology, perceiving it as a useful tool that may enhance conventional teaching techniques and promote more inclusive, dynamic learning settings. Digital teachers can play a complementary role in the classroom and have irreplaceable advantages of accessibility and equity.

In the following sections, we first thoroughly examine the technologies in educational settings by exploring pertinent literature on pedagogical strategies and the use of "digital humans". After providing this basic perspective, we outline our AI lecturers' technological implementation and design components. Our methodology section describes our mixed-methods approach, which includes quantitative surveys with 26 students to define general perception and rankings, as well as qualitative semi-structured interviews with 15 volunteer participants to provide deeper insights into their experiences. The quantitative data supports the later in-depth interview. In the findings and discussion, we explore the consequences of our research, providing design principles for the next efforts using digital teachers and imagining how technology may improve access to education, especially for marginalized communities and MOOC platforms. We aim to stimulate more research and innovation in the field of digital education, as this is the first study of its kind on AI digital instructors in higher education with high media coverage.

## 2 BACKGROUND AND RELATED WORK

### 2.1 Existing Pedagogical Practices and Transformation

A large body of research on successful teaching and educational effectiveness emphasizes the necessity for a change from teacher-centered, authoritarian methods. Studies conducted in the context of higher education revealed that students are losing faith in teacher-focused pedagogy [7] and are becoming less interested in learning and teaching paradigms that they believe to be antiquated [8, 25]. Discussion was in the midst of the commonsense teaching (derived from tradition and lore) and trained or expert teaching (derived from empirical models and knowledge) [58]. Much research highlights that external factors, such as teachers' attitudes and perspectives, are crucial to teaching effectiveness [9, 23]. As such, modern educational concepts no longer regard students as passive recipients of information; but have shifted to an inclusive model, in which students play an active role in education and have the ability to influence both procedures and content.

Earlier educational approaches, such as constructivist, Vygotsky [64] and Persson [58], encourage students to investigate, experiment, and ask questions, are the foundation of the current paradigm shift. These transformative and constructive approaches focus on active student participation, social and culture enactment, and develop a cordial teacher-student relationship [18, 25, 35]. Instead of being lectured on abstract concepts, instructors encourage students to construct their own knowledge via experiences and activities. As such, learning becomes an active process. In their ideology, a traditional teacher delivers didactic lectures that cover the subject matter, whereas the role of instructor shifts to that of a facilitator, assisting students develop their own understanding of the content.

Constructivist assessment strategies include oral discussions, hands-on activities, KWL(H) chart, and pre-testing [11, 36, 47] . However, opponents argue that there is insufficient empirical evidence to support the core assertion and effectiveness of constructivism. Scholars [44, 48, 61, 63] reckon this approach is not friendly to novices and describe constructivist teaching methods as problematic "unguided methods of instruction". Inspired by constructivism, theorists developed more well-known learning methodologies, such as discovery, problem-based, and inquiry-based learning, laying the foundation for modern pedagogical practices.

Contemporary pedagogical practices are increasingly embracing a more student-centered approach. Some researchers have found that the real merits of teachers' teaching practice depend primarily on their embrace of "humanistic" pedagogy rather than being rooted in subject matter expertise and solid pedagogy. [9, 23, 24].

One-to-one pedagogy, as one of the "humanistic" and "customized" pedagogy, is being criticized for its workability, applicability, and transferability [12, 49, 58, 59]. Scholars claimed that it has detrimental long-term impacts on students as learners, resulting in reliance, less prospects for employment, and disappointment [12, 13]. Meanwhile, some believe this approach is indeed potential for a productive mentoring environment - not least of which is that it might help students make the connections between what they are studying at the conservatoire and career paths. Despite these discrepancies, this prevalent practice, particularly in tertiary vocal and instrumental tuition, is seen as essential for student learning and development. While one-to-one pedagogy foregrounds the customization and conservatory roots, collaborative learning stems from process theories of rhetoric and composition, emphasizing engagement and collaboration.

Collaborative pedagogy is effective to scaffold learning and promote critical and reflective thinking in students via discourse [26]. It has been commonly used in higher education classrooms through the emergence of group assignments and web-based learning management systems (LMS). Skeptics criticize its methodological juxtaposition, arguing that it values group work over individuals, in stark contrast to the university's values [10, 33].

These major pedagogical practices are being condemned in the vicinity of their learning and teaching effectiveness. There is no 'perfect' way to reach consensus. A more holistic approach should therefore be akin to a tapestry, woven from traditional and innovative pedagogical practices [5], to enhance the overall educational experience. With this said, understanding modern teaching approaches with technological advancement is crucial.

## 2.2 Learning with Technologies

The integration of technology has transformed the traditional teaching paradigm. Technology has left its indelible mark on education, improving classroom teaching effectiveness by promoting student-centered learning and teacher-student relationships [62]. Digital tools and computer-based learning environments has opened up new avenues for instruction and engagement, promoting collaboration and problem-solving skills [32, 40]. Labadi's team concluded that using technology outweighs traditional instructional methods via comparing students' academic performance and understanding level in statistics courses [6]. From asynchronous text-based CMC to the learning management system (LMS), they enhance education's accessibility, personalization, and interactivity [39, 53].

The recent shift to online learning, necessitated by the COVID-19 pandemic, has further underscored the pivotal role of technology in education [51]. With the advent of many large language models such as the GPT family [28, 65] and Metaverse [20, 21, 29, 52], educational technology is starting another chapter. Learning has thenceforth evolved in tandem with these developments.

The powerful GPT family models accelerate various sectors. Dwivedi et al. investigated these ramifications, looking at the potential and difficulties that emerge from ChatGPT and other generative AI (GenAI) technologies in the fields of business, education, and society [22]. Researchers explore a variety of applications and scenarios that can emerge GenAI into education, including educational games, chatbots, AI-generated synthetic learning videos, and software development education [28, 30].

While Tzirides and the team reviewed that chatbots responding from large language models (C-LLM) can potentially enhance education by reviewing and assessing complex student work, its limitations remain bound to language corpora and binary notation [54]. Mogavi et al. further delved into how ChatGPT has been used in education [28], revealing the fact that early adopters have mixed perceptions on how GPT was used in educational settings. As one would reasonably expect, other scholars have also expressed concerns over the ethical challenges of using GenAI tools in education, calling for the needs of effective guidelines and policy making [14, 16, 57, 60]. Thence, these considerations should be carefully examined and evaluated in every GenAI educational application, as the digital teachers we put forth in this work.

On the other hand, immersive technology is becoming more impactful with the arrival of the Metaverse new era [21]. Studies examine the effectiveness and students' feedback on using virtual reality (VR) technology in learning various subjects, such as forensic science chemistry education [41], adult learning [55], clinical sciences simulation and training [42]. Immersive technologies, particularly VR, can enhance student learning outcomes under the replica of real-life scenarios [17]. More applications are thenceforth developed for immersive education, including a multi-device immersive learning environment via CHIC Immersive Bubble Chart [56], and ImmerTai that allows learning of Chinese Taichi in VR [15]. This transfer learning phenomenon has been witnessed in many contexts across different target user groups with different intended learning outcomes (ILOs).

Digital human, also known as avatar and agent, is one of the core elements in immersive technologies [46], and is commonly put in immersive learning applications. Virtual Objective Structured Clinical Examination (VOSCE) is one of the early attempt to put virtual human in the educational sector [38]. They tried to validate the use of virtual humans for interpersonal skills education through comparing real and virtual human interaction.

Later studies further supported the use of digital humans in education. Zhou et al. emphasize that AI digital humans can enhance English oral teaching and foster novel learner experience, creating inclusive educational environments. Wu et al. [67] discuss a less direct application, Mr. Brick, where children's actions in the physical world affect the mixed-reality environment. It helps children adapt to remote collaboration and simulates human interaction digitally. Mogavi et al. [31] mention that user avatars in Stack Exchange's Winter Bash act like digital humans, promoting a sense of community and achievement, which are crucial for motivational learning environments. Zhang et al. [68] further analyze digital avatars and their role in popularizing knowledge, highlighting how these virtual representations can positively influence learning effect, emotional experience and user engagement dimensions.

It is conceivable to conclude advanced technology is useful for customizing learning resources. Existing research proved the effective outcomes for immersive education in STEM [45] and foreshadowed the future direction beyond scientific subjects. This leaves a perceivable question: how should practitioners leverage existing technologies to create better teaching and learning experience?

## 3 METHOD

This section describes our participants' recruitment process, class and study arrangement, participants, data collection, and analysis.

### 3.1 Participant Recruitment

*3.1.1 Recruitment Process.* Before we began our courses and research, we were approved by our university's Institutional Review Board (IRB) to ensure that our work complied with all ethical guidelines and procedures pertaining to research involving human subjects. In view of the special context of our study, the research methods should be grounded in the facticity and applicability of actual higher-education scenarios. Following IRB approval, the study ran through a semester-long postgraduate course, beginning with our university's typical course registration process using convenience sampling techniques.

The advantages of involving student participants lie primarily in its relevance and authenticity. Students are the primary stakeholders in the education system, so their involvement ensures that the research is directly relevant to the target population. The core objectives of our study are to support student learning, holistic development, and potential advancements or future directions for digital teachers. Thus, students' insightful comments and feedback on digital teachers' viability, acceptability, and practical implementation align with our goals.

Meanwhile, challenges with this approach are generally due to its voluntary nature. Participating in research activities may disrupt students' normal learning and classroom experiences. Therefore, researchers employed regular refinement through rigorous pilot testing, providing transparent communication and expectation setting to minimize bias and disruption to the student's regular learning experiences. Most importantly, all participants in this study were voluntary and could withdraw at any time during any of the ten weeks. They were free to decline at any point or to not respond to any question for any reason.

To better recruit our participants, we distributed a course teaser on YouTube and Bilibili and later promoted it via school email. The course teaser demonstrated the use of digital teachers, course descriptions, Course Intended Learning Outcomes (CILOs), time, and venue. It also detailed the research's objectives to ensure its transparency.

*3.1.2 Recruitment Considerations.* Participants were required to be current students and on a first-come, first-served basis. Since this cross-campus course targeted an interdisciplinary background, we welcomed all students from different faculties. As the number of head-mounted displays (HMDs) is limited in the two campuses, the recruitment quota varied. The number and information about the participants will be further reported in 4.3 Participants.

Measures were implemented to ensure that participants were well-informed, comfortable, and able to voluntarily engage in the study while considering their individual experiences and expectations. First, students could withdraw from the study during the add/drop period from January 31st to March 1st to uphold voluntary participation. Second, a pilot lesson was organized in the first week to familiarize students with class arrangements, rubrics, curriculum, and grading criteria. Third, participants completed a Qualtrics formative consent form before enrollment. The online consent outlines the study's objectives, confidentiality policies, and demographic questions, including gender, age, and contact details. The form also assessed prior technological experiences with virtual reality (VR), including usage duration, expertise level, and familiarity with digital humans and technologies, as well as participants' expectations and concerns regarding digital teachers.

To ensure fairness, student mobility was acknowledged, allowing participants to withdraw or add classes as needed, with absences for illness or personal leave deemed acceptable. Consequently, the number of participants in each lecture varied, making post-survey findings primarily a reference for comparing digital teachers' popularity and informing interview questions. Ultimately, our research is grounded in in-depth semi-structured interviews, providing insights into participant experiences with digital educators.

### 3.2 Class and Study Arrangement

To ensure the quality of teaching in the postgraduate course CMAA5022 / EMIA6500F Social Media for Creatives, we conducted weekly rehearsals involving two research assistants, four teaching assistants, and researchers. Each rehearsal, lasting 60-90 minutes, included an overall rundown of the class, testing of digital teachers, proofreading of teaching materials, and equipment checks, such as meeting connections across campuses and installation of head-mounted displays (HMDs).

The course adopted a hybrid teaching approach that integrated digital and human lecturers. This interdisciplinary curriculum invited students to explore social media platforms, content creation strategies, and generative AI technologies (see figure 2). Key topics included digital storytelling, gamification, immersive technologies, and relevant research skills, emphasizing exploration and creativity. The use of digital teachers complemented the course's context, providing multi-accessibility across various formats. Beyond that, the curriculum and teaching methods were approved by senior academic staff, ensuring alignment with educational standards.

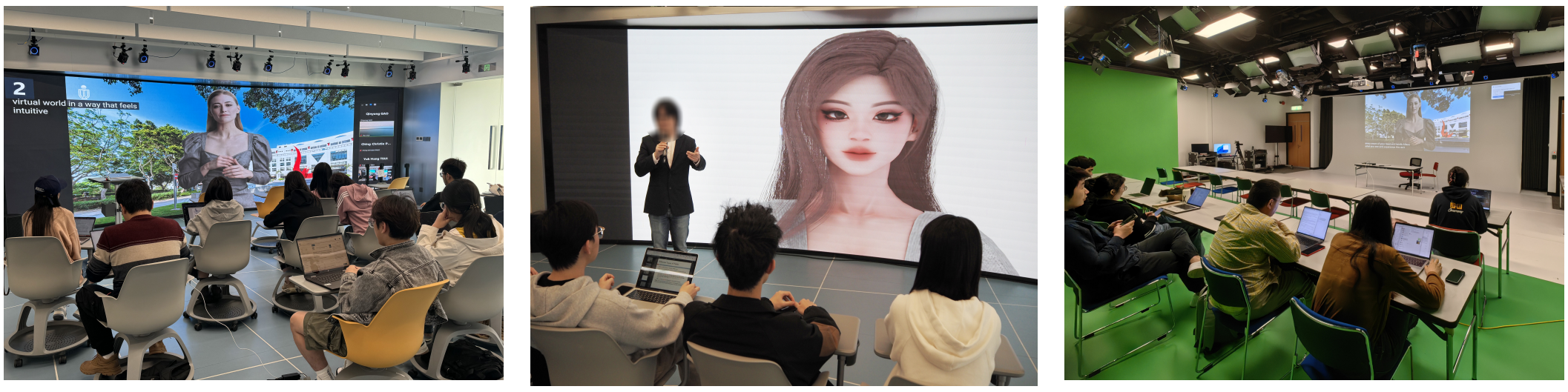

(a) Course photo in Guangzhou campus.

(b) Course leader was introducing our digital teacher.

(c) Course photo in Hong Kong campus.

Fig. 2. This is a cross-campus graduate-level course held simultaneously on both campuses every week. Digital teachers met with students in different formats.

This cross-campus course, scheduled for Spring 2024, consists of three hours of instruction per week. Approximately 30 minutes of each lecture is dedicated to explanatory content covering social media terminology, while 60 minutes focuses on case studies and theoretical knowledge. The remaining time includes paper presentations, in-class discussions, and hands-on tutorials led by teaching assistants, interspersed with breaks of 5-15 minutes.

Given the collaborative nature of the course, students and teaching teams were distributed across both campuses, with all lectures recorded and live-streamed via Zoom. Following discussions with senior academic staff and feedback from an internal pilot study, it was decided that digital teachers would cover one-fourth of the teaching materials per class, facilitating a balance of interactivity. The study was conducted over ten weekly lectures, from January 23 to April 16, 2024, with each session lasting 30 minutes and featuring different digital teaching styles.

Inspired by the MUD card method [27, 34, 50], participants provided feedback via Qualtrics after each section, ensuring the voluntary nature of participation through consent notices. Surveys, lasting three to six minutes, included Likert scale items, multiple-choice questions, and short-answer prompts regarding students' preferences and ratings of digital teachers. In the eleventh lecture, students ranked the digital teachers and had the option to participate in voluntary in-depth interviews, emphasizing the discretionary nature of their involvement.

Upon completing the 10-week study, an overall evaluation survey was conducted in Week 11, where participants ranked the digital teachers they had interacted with and were invited to volunteer for in-depth follow-up interviews, fostering a participatory research environment. This multi-pronged approach, combining quantitative rankings with qualitative insights, aimed to deepen our understanding of participants' experiences and perceptions regarding the effectiveness, usability, and overall value of the digital teacher applications examined.

### 3.3 Participants

Our study involved 27 students, of whom 2 withdrew and 1 later rejoined, resulting in a final participant count of 26, designated as P1 through P26. The cohort consisted of 12 females, 13 males, and 1 participant who preferred not to disclose gender. Participants primarily came from diverse professional backgrounds, with 9 in computer science, machine learning, engineering, and science-related fields, 6 in art and design, and the remainder enrolled in interdisciplinary programs. Participants self-identified their expertise, with 13 labeling themselves as "Technician/Engineer" (5 females) and 10 as "Designer/Artist" (4 females), while 3 opted not to disclose their professional backgrounds. Geographically, the majority of participants hailed from Mainland China (25 participants, 12 females), with 1 participant from Egypt. The cohort included 25 research postgraduates and 1 undergraduate, reflecting the university's flexible course registration policy. Participants' ages ranged from 18 to 45, with an average age of 26 years (SD: 5.21), distributed across three cohorts: 13 participants aged 18-25, 12 aged 26-35, and 1 aged 36-45.

Participants were queried about their familiarity with digital humans and immersive technologies, which are pertinent to our study. Responses varied significantly. Regarding VR experience, 5 participants reported no relevant experience, 13 had less than six months, 2 had six months to a year, 3 had one to two years, 1 had two to five years, and 2 had over five years. Frequency of engagement also varied, with 9 participants rarely or never using VR, while others reported engaging from once a month to daily. Most participants identified as "beginners" (17), with 7 considering themselves "intermediate" and 2 as "advanced" users.

Prior to the study, all participants were familiar with digital human applications, mainly through entertainment. Notable virtual icons mentioned included Luo Tianyi (18 mentions), Barbie (8), Kizuna AI (3), AYAYI (3), and others, indicating a strong recognition of digital humans in social media contexts.

The above summarizes demographic data for students in the class. As mentioned, our study strongly counts on these interviews. The following section illustrates the data collection process and highlights the details of our in-depth interview.

### 3.4 Data Collection

The purpose of the interviews was fourfold: (i) to investigate how the variation in digital teacher styles and features impacted students' perceptions and experiences; (ii) to examine how students' background affected their acceptance and preference towards digital teachers; (iii) to explore the impact of the presence of digital teachers on teaching methodologies and pedagogical practices in higher education; and (iv) to invite participants designing their ideal digital teacher as inspired by the typical style of participatory workshop.

Among the 26 participants in class, 15 volunteered for interviews (P1-P15), comprising 8 males and 7 females, with 4 identifying as designers/artists and 11 as technicians/engineers. Appendix A 6 depicts the character design worksheet. Table 1 shows the participants' demographic and their preferences collected from the post-lesson survey.

Table 1. An overview of the semi-structured interview participants' demographics and preference on digital teachers after the 10 weeks experiences

| ID | G | Platform | Expertise | Age Group | Best Teacher | Most Valued Feature | Interest in Digital Human Application(s) |
|---|---|---|---|---|---|---|---|
| 1 | F | F2F | TE | 18-25 | Einstein | Format | Yes, occasionally |
| 2 | F | F2F | TE | 26-35 | Fiona | Style and Design | Yes, occasionally |
| 3 | F | F2F | DA | 18-25 | Fiona | Style and Design | No, but I'm interested in trying it |
| 4 | F | F2F | DA | 18-25 | Camilla | Non-verbal Cue | Yes, occasionally |
| 5 | F | F2F | TE | 18-25 | Fiona | Non-verbal Cue | Yes, occasionally |
| 6 | M | F2F | TE | 18-25 | Einstein | Format | No, but I'm interested in trying it |
| 7 | F | F2F | DA | 26-35 | Fiona | Style and Design | No, but I'm interested in trying it |
| 8 | M | F2F | TE | 26-35 | Fiona | Non-verbal Cue | Yes, frequently |
| 9 | M | F2F | DA | 26-35 | Fiona | Non-verbal Cue | Yes, frequently |
| 10 | M | F2F | TE | 18-25 | Ben | Format | Yes, occasionally |
| 11 | M | F2F | TE | 26-35 | Ben | TQC | Yes, frequently |
| 12 | M | F2F | TE | 18-25 | Idol Aria | Format | Yes, occasionally |
| 13 | M | Zoom | TE | 36-45 | Ben | Style and Design | Yes, occasionally |
| 14 | F | Zoom | TE | 26-35 | Gabriela | Format | No, but I'm interested in trying it |
| 15 | M | F2F | TE | 18-25 | Einstein | TQC | Yes, occasionally |

Our flexible interview methodology ensures the open and inclusive expression of ideas. For individuals who preferred not to conduct in-person interviews, we accommodated them with online interviews (n=2) via Zoom. The interviews with other participants were mainly conducted in the campus and arranged in accordance to their schedules (n=13). Furthermore, Interviews were performed in the participants' native tongue, mainly Mandarin (n=14) and English (n=1) to ensure their comfort, allowing for free and unrestricted idea expression.

The data collection process for the interviews was from the 17th of April 2024 to the 31st of May 2024. Each interview lasted between 50 and 75 minutes, typically in a meeting room, with posters and all digital teachers' screenshots showcased to recall the learning experiences (see figure 3). After the brief warmup and overview, each interview question was subject to individuals' answers according to the former survey results under the same structure. Interviews were also audio-recorded with participants' permission. The worksheets were collected under participants' consent to formulate the ideal digital teachers and their future vision (see figure 3).

To ensure the accuracy of the transcripts and translations, each recording was first transcribed using iFlytek transcription tool, and manually validated verbatim by at least two researchers. Translation was performed among three researchers parallelly. Note that the three researchers are native Chinese speakers, of which we kept most of the language accurate. Researchers manually rewrote, rephrased, and adjusted the transcription errors and incoherent sentences. By comparing the three versions of translation, researchers discussed and agreed to the final translation version. To further validate the accuracy and maintain the essence of the interviews, researchers used ChatGPT 4 to perform a polished and extracted version of the text. This simplified version mainly removes some repetitive and redundant parts, as well as some invalid information such as oral thinking. The final analysis was built after the mutual agreement on the lexical choice, comparison between the simplified and original text, and consideration over the participants' cultural and age with iteration. Overall, the data collection process was prepared under systematic and detailed procedures.

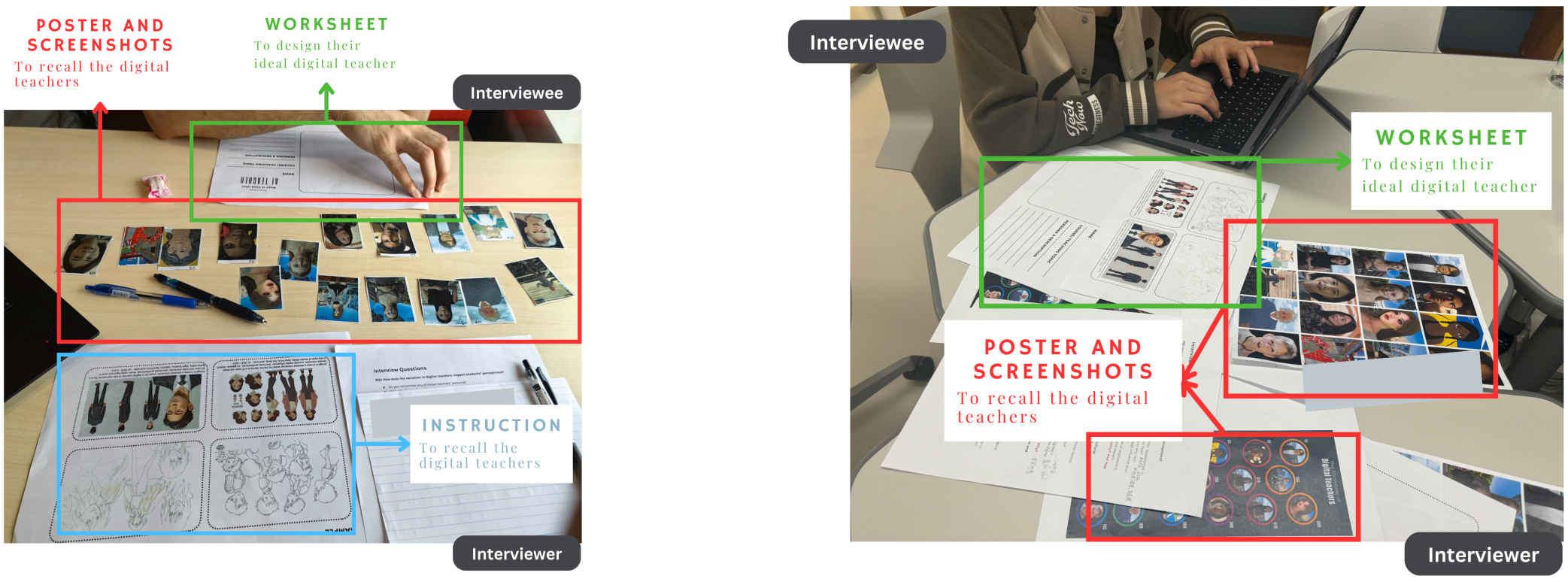

Fig. 3. Interview and its setup in the study.

### 3.5 Data Analysis

The study employed a mixed-methods approach, combining quantitative and qualitative data collection and analysis. The quantitative component involved a survey of 26 students to assess their rankings and ratings of the digital teachers. The qualitative component consisted of semi-structured interviews with 15 volunteer participants from the course. While the data collected from each survey was tabulated and compared to identify patterns and trends in the students' perceptions of different digital teachers, our data analysis focuses on the data collected from the semi-structured interviews.

The semi-structured interviews were transcribed and subjected to inductive analysis. The researchers, comprising two postgraduate students and a postdoctoral research fellow, held weekly meetings via Zoom to perform the systematic qualitative data analysis using the collaborative whiteboard tool named Miro [1]. The analysis involved creating affinity diagrams and engaging in in-group discussions to identify emerging themes, patterns of perception, and primary concepts in potential improvement or future direction. Figure 4 shows the instance of affinity diagramming in our Miro board.

Refinements were made through the two-month discussion iteration, enabling us to fine-tune the material. Specifically, we improved the drafted groupings and naming. For example, "realism and credibility" were originally an independent theme for key factors in RQ1. It was yet found that many shared features overlapping with other groups. After thoughtful discussion, three researchers agreed to regroup the content, integrating the concepts into other main themes. Another significant refinement was renaming the defined pattern. For example, "Enhancing the Interactive Elements" was renamed to a more matching appellation called "Situation-Aware Interactivity" in line with the corpus and context.

[1]https://miro.com/

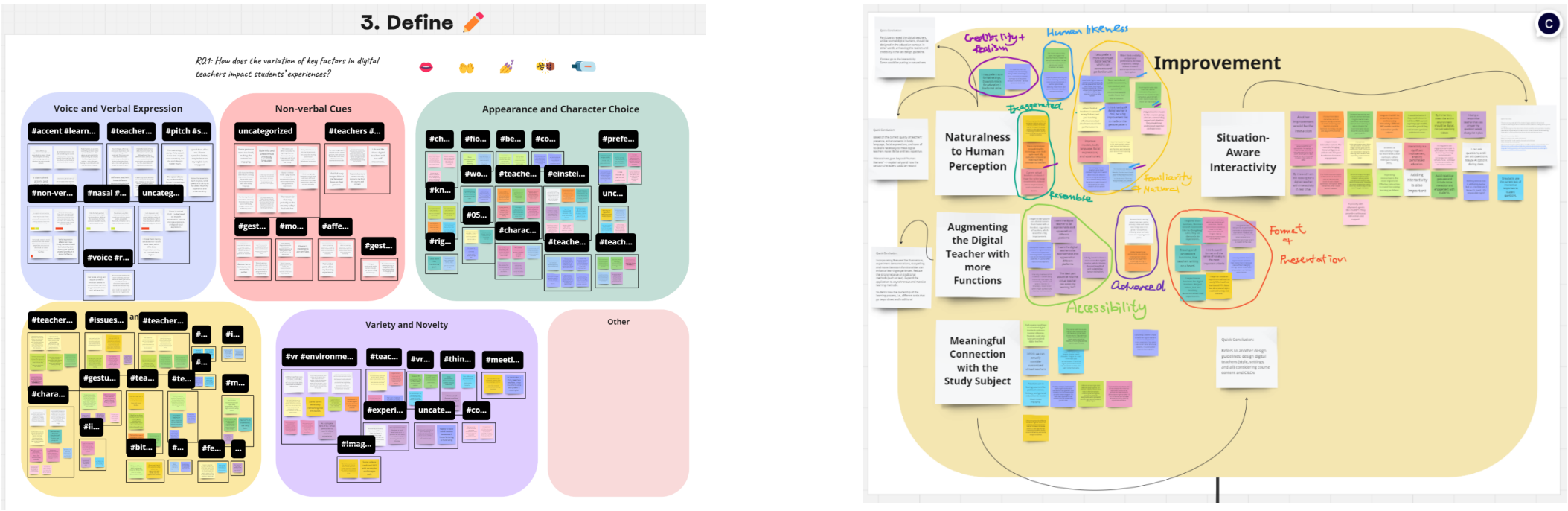

Fig. 4. Affinity diagramming session in Miro was used in our study.

After the rounds of intensive discussion and revisiting the interview materials, we finally reached the consensus of (RQ1) four key factors in digital teachers that impact students' experiences as (I) *voice and verbal expression*, (II) *non-verbal cues*, (III) *appearance and character choice*, and (IV) *variety and novelty*; and (RQ2) four themes on the suggested improvement and future direction for digital teachers as (I) *situation-aware interactivity*, (II) *naturalness to human perception*, (III) *augmenting the digital teacher with more functions*, and (IV) *meaningful connection with the study subject*. Overall, we prioritized the authenticity of the findings. That said, our analysis comports with the collected data. In the next section and its subsections, we will expound upon these themes that we have identified.

### 3.6 Positionality Statement

Our research team, comprising postgraduate researchers, a faculty member from East Asia, and a postdoctoral scholar from the Middle East, recognizes the importance of addressing potential biases to ensure fairness in this study. Our shared cultural background with the majority of participants (N=26) provides valuable insights into the local context influencing students' perceptions of digital teachers. However, this proximity may also introduce unintended biases in framing research questions and interpreting data.

To mitigate these risks, we adopted a systematic approach emphasizing objectivity and transparency through iterative processes, including regular peer reviews and collaborative data analysis. The course instructor validated the content of the digital teachers, whereas a non-East-Asian postdoctoral researcher monitored data collection to minimize biases. Additionally, the involvement of teaching assistants enhances our understanding of teaching dynamics, allowing for better contextualization of student feedback.

By being transparent about our positionality and the measures taken to address potential biases, we aim to ensure a rigorous and fair representation of students' diverse perspectives on the integration of digital teachers in higher education, establishing a foundational sample for future implementations.

## 4 FINDINGS

This section describes the details of our participants' recruitment process, class and study arrangement, data collection, and analysis.

## 4.1 RQ1: How does the variation of key factors in digital teachers impact students' experiences?

RQ1 investigates the main elements of digital teachers that influence students' learning experiences. Table **??** in the Appendix section shows four key aspects of these elements that were identified through interviews and summarized from the data analysis. The following subsections will elaborate on how these factors in digital teachers can impact students' learning experiences.

*4.1.1 Appearance and Character Choice.* The appearance of digital teachers created a crucial first visual impression, shaping students' overall perception of the learning experience. Student feedback can be categorized into three main areas. First, the selection of characters for digital teachers influenced their perceived credibility. Second, the rendering style played a critical role; and third, attributes such as gender, age, and ethnicity also impacted students' acceptance of the digital teacher.

Students' perceptions of digital teachers were significantly influenced by digital teachers' appearances and the chosen persona that represent them. In other words, students were attentive to who the digital teachers are. Initial recognition and impression of the chosen characters impacted the credibility assigned to the teachers. If the characters were the digital versions of real-world figures or authorities previously known, it considerably enhanced its perceived credibility. For instance, P6 praised the digital version of Einstein, stating, "*Being a science major at the postgraduate level, I feel the class gains more substance when Einstein teachesxsl.*" Another student, P5, explained the rationale behind choosing recognizable figures like Einstein or the digital version of their course professor: "*For well-known figures like Einstein or our professor, there's a sense of familiarity,*" while yet another (P15) remarked, "*This is why I prefer the digital version of our professor (comparing to other digital teachers). I am more familiar with him since I know him well.*", and P5's statement proved this inference from another direction by expressing his indifferent feeling towards the digital teachers that were not his pre-known characters: "*...but honestly, after the classes, I feel like I am indifferent to digital teachers (that I did not know), since I am not familiar with him/her*".

Regarding how the the rendering styles of digital teachers influenced the students' learning experience, students had their own preference of whether they subjectively liked or disliked either style. But when judging the two styles from a more objective perspective, it could be summarized as that they prefer realistic style for real-human like characters, and photo-realistic/cartoonish style for comic characters. P5 said, "realistic ones feel more like real teachers, while two-dimensional ones feel like anime characters teaching, which is also appealing and more amiable". It can be inferred that students liked the realistic for the more professional scholarly image. Although this inference might not carry much meaning, P12's put may help to explain why so: "*It (the mixed of realistic and comic) is like having a very realistic doll, and that can be frightening*". P3 also addressed the same idea, stating, "*But overall, they should either be very lifelike or very natural. If it's a realistic style, it should be as natural as possible; if it's cartoonish, it should be more animated*".

Furthermore, attributes, such as gender, age, and ethnicity, did affect some students' learning experiences with digital teachers, while others did not consider these factors significant. Among the students who thought these attributes influenced their learning experience, student P14 expressed, "*I think age, gender, and appearance are quite important for my learning experience*". Some students found the similarities the characters shared with them to be important. For example, student P7, a female student, remarked on her attitude towards digital teacher Fiona: "*She was also an Asian woman, which felt familiar and approachable*". Meanwhile, other students' reasons were based on their perception of the attributes of a typical post-graduate teacher. Student P15 believed that alignment with the typical image of a senior professor increased credibility, commenting, "*While I think credibility is the most important, I always believe a mature woman professor is the best option*".

He also noted, "*Because of my [cultural] background, I prefer female. People at my age and from my time think women are more gentle and tender*".

*4.1.2 Non-verbal Cues.* Most students valued non-verbal cues of digital teachers' class performance when evaluating their learning experience. The mentioned non-verbal cues include the overall body movements, and the specific body language like postures and gestures of digital teachers. The general movements of digital teachers formed the basis for students' evaluations. As one student (P9) said "*I think having body language is good. Some [digital teachers] only have stiff facial movements and single expressions, which are not very engaging*", the inclusion of body language was important as it made the teacher's behavior more human-like, thereby making students feel more engaged and focused in the class.

The students also commented on the qualities of digital teachers' movements that have influenced the qualities of their learning experience. First, it can be referred that naturalness and smoothness are important for their learning quality. Student P9 commented that "I do not like those digital teachers with too stiff movements," and "Eleanor's movements are very fake". Student P8 also supported this view, giving credit to the digital teachers whose movements are smooth: "*I chose/[ranked] Gabriella second because her movements are smooth.*" Students P7 and P5 also emphasized the importance of natural movements, explaining the reason behind this: "*Unnatural movements make me focus on them [the digital teachers] rather than the course content*" (P7); "*My focus is more on whether their lip movements are natural, which affects my attention to the content*" (P5). It can be inferred from both of the reflections that natural movements do not distract students, allowing them to focus more on the learning materials.

For body postures and gestures, students suggested an even more ideal scenario where the digital teachers' movements include variations, as inferred from student P4's negative feedback regarding repetitive body gestures of digital teachers: "*Some gestures were too fixed, making the content less engaging*". This was also supported by positive feedback about the lack of repetition in some digital teachers' gestures: "*I chose/[rank] Gabriella second because her facial expressions and micro-movements are not repetitive, which is good*" (P8). A student's feedback and reflection on certain digital teachers can explain the reason: "*A repeated gesture pattern actually distracts me from the course and content*" (P15). Therefore, it can be inferred that, connecting to the earlier discussion on naturalness, repeated gestures make the digital teachers appear more unnatural and fake, which distracts students. Lastly, if the gestures are consonant and consistent with the verbal delivery, it would be an added bonus, as student P1 suggested: "*The main thing is when a digital teacher added gestures, their gestures moved with their tone, which I found very novel and memorable.*"

*4.1.3 Voice and Verbal Expression.* The voice and verbal expression carried the content of the course. Only a few students (N=2) reflected that they did not pay much attention to the voice and verbal expressions and thought those did not matter to the learning experience. Most students had different comments on how the voice and verbal expressions influenced their learning experience and acceptance of the class materials. For example, student P2 expressed her own preference: "*In a slightly serious teaching environment, we usually think a more mature voice is more convincing. But in fact, many people chose the sweeter voice, possibly because it feels more relatable and friendly.*" Student P15 also expressed his own preference for a more mature voice. However, no matter what students personal preference were, regarding the pitch and tone of digital teachers, student suggested it should follow a principle which is the naturalness, so students would comprehended the class materials with more attention and were not distracted by the digital teacher's voice awkwardness. As student P10 addressed, "*Voice affects my experience. Different teachers have different tones and pitches, which is fine as long as it sounds natural". So, students thought the overall naturalness was more important than the specific pitch or tone of the digital teachers*".

Unlike the differences in personal preferences for pitch and tone, students had quite consistent opinions regarding whether digital teachers had accents and how accents influenced their learning experience. They thought accents were normal, and having an accent made the digital teacher more authentic to them. Students preferred having accents in their classes with digital teachers. As student P14 noted, "*I don't think accent and voice impact my learning experience*" , and student P13 remarked, "*To be honest, an accent is something we (non-native speakers) all have ... But when we get used to this kind of thing, we actually feel like, hey, it does't matter... or in other way, having an accent makes me feel like this is a human*". Similarly, this logic was also applicable to the fluency of the verbal cues of digital teachers. Students endorsed digital teachers that did not have perfect verbal fluency. As student P15's statement inferred, "*A too fluent English makes me realize it is a fake character. No one can speak without stammer and mistake. Some digital teachers speak so fast, so fluent without stuttering and pause, this affects my learning, especially I am not a native speaker*".

Based on the analysis of students' feedback, the speaking pace and speed could be categorized together. The reason is that both influenced students' understanding of the course materials delivered by the digital teachers. Having a proper pace of speech helps students understand the content, as pauses and sentence segmentation help them follow and comprehend key points. Student P6 criticized a digital teacher whose speed did not include enough pauses: "*Lack of pauses and discussions on specific points felt like a waste of time. Sometimes, half an hour would pass without understanding the content*". He also affirmed the importance of having pauses to maintain the overall rhythm of the speech, even without mentioning content comprehension: "*The main thing is how much I can understand. Some teachers, although I don't fully understand them, I like listening to them because they pause and don't speak in long, continuous segments*".

As for the speed of the speeches, students consistently agreed that speed influenced the understanding of speech content. Especially for students who were more concerned with their English comprehension skills, the speed was important, and slower speeches would be more listener-friendly for them. These inferences can be drawn from student P6 and P9's statements: "*Speed does affect me. Slower speech is better, maybe because my English isn't very good*", and "*The speed affects my understanding. Slower is sometimes better.*". Student P3 also emphasized that slower speed was important for clear speeches: "*The main thing is clarity. If the speech is too fast, I might miss something*".

*4.1.4 Variety and Novelty.* On average, students cherished the different forms of the digital teachers' presentations. There are digital teacher presentation formats in video and VR. Except for two sessions presented in VR, the rest of the sessions were presented in the video format. Students pointed out aspects that affected their learning experiences in this presentation format. They noted that it would be better if the digital teachers' lectures were supplemented with visual aids like motion graphics and videos corresponding to the teaching content. This method not only explained what the digital teachers were talking about but also stimulated students' learning interests. This finding can be inferred from students P4 and P15's statements: "*Digital teachers can help in the classroom because Some videos were well-edited, combining images and dynamic explanations, making long classes less tiring*"(P4); "*The video interaction and content with the avatar imagery make the learning interesting*"(P15).

Students' evaluations of the VR presentation format were quite positive, and there are different reasons for that. First, the digital teachers in VR format gave students a deeper impression, and they ranked digital teachers in VR higher. As student P1 stated, "*Idol Aria's class format is very innovative. It is a VR format. This exploratory teaching method left a deep impression on me, so it is ranked higher.*" This indicates that the digital teachers' teaching in VR format might help students memorize more course content covered in the class, thus resulting in better learning outcomes.

As student P10 mentioned, "*Digital teachers increase my focus [concentration level] in class. In VR, I would look at exhibits more, leaving a deeper impression.*" Similar feedback can be found in student P7's evaluation: "*I remember more content from Einstein's lessons in VR.*"

Additionally, digital teachers presenting in the VR format could promote students' learning interests because of the immersive learning experience it provides. Student P11 stated, "*VR has advantages. The environment is more immersive, which helps engage my learning interest, especially with traditional, scripted content.*" Similar feedback from another student indicated that VR increased their interest, motivation, and engagement in the class, adding that VR promotes a more active attitude towards learning the class content: "*VR increases my engagement, and I'm more willing to explore new content.*" Moreover, VR seemed to offer another way of learning, empowering students with the ability to control the class and manage their learning experience themselves. As student P11 put it, "*In one class, we self-studied with VR, exploring independently, which improved self-learning and focus.*"

More specifically, the freedom to move, the rendering, and the functionality in the VR presentation format provide students with abilities they do not usually have in physical classes. Students P6 and P1 endorsed the ability to move so they could adjust their distance to the slides in the front of the virtual classroom. Student P6 remarked, "*3D might be more helpful. The PPT in VR is very large, which we can't usually see in such detail during regular classes.*" Similarly, student P1 noted, "*In VR, the environment setup, with blue skies and white clouds, allows me to stand by the classroom window and look at the clouds. I felt that the environment had a significant impact on my learning, so if VR could provide this kind of pleasant environment, it would be beneficial for my studies.*" They also appreciated the adjustable seating, with P1 saying, "*In VR, I usually look at the notes closely, and during the class, I want to see them from a distance, so the adjustable seat was very convenient. When I wanted to see the notes, I could just stand under them and look at them closely, which was great for my vision.*" The environment rendering in the VR format influenced the students' learning as it created an atmosphere conducive to learning.

However, the VR format does have its limitations due to the device properties, such as the motion sickness it could bring to users. As student P15 noted, "*VR teaching is mentally demanding for me. I mean... personally, I can only stay there for 10 minutes concentrating...*", It may be necessary to consider constraining the session length for digital teachers' lectures to a proper period to avoid uncomfortable experiences for students.

### 4.2 RQ2: According to students, what are the potential areas for improvement in the implementation of digital teachers, and how do they view the future of digital teachers in education?

The succeeding parts describe how the digital teachers can be potentially improved according to the students. Table **??** in Appendix shows the main themes of the suggestions, their frequency that was being mentioned by interviewees, and the key takeaway. We will discuss deeper the four defined themes of improvements in the following sections.

*4.2.1 Situation-Aware Interactivity.* Most participants expressed that "situation-aware interactivity" was in strong need of improvement. Note that the existing digital teachers, both in video and VR formats, are not responsive. P8 carefully recapped his learning experience in class, saying, "*having interactivity [in digital teachers] would be a significant improvement, enabling personalized education in the way that can answer individuals' questions.*" When asked about the type and definition of interaction, he further elaborated, "*AI should be able to interact with students and answer questions… I mean… with interaction, all other issues (the styles and aesthetic design) can be resolved easily. Interaction is the most influential factor in communication.*" Another participant, P6, also inspired by the class, captured, "*currently, it (the digital teacher's lesson) is just a one-way communication... she speaks in the video and we listen to her,*" and stressed the need of having interactive teachers, by saying, "*more interaction, such as teacher asking questions, then answered by student, and teacher providing the feedback in a loop, should be a normal way of communication we typically have in reality.*" He further mentioned his experience in using AI Chatbot, suggesting, "*I am quite used to ChatGPT so I believe integrating [ChatGPT] for answering questions can increase interactivity and facilitate our learning process. I can memorize easier.*" He concluded this suggestion with his experience and said, "*different GPT models could be trained for each specific subject.*"

P1 expressed her strong desire in having interactive digital teachers by comparing traditional teaching and active learning approach, saying, "*in terms of interactivity, I am actually looking for more interactions rather than just reading PPTs (slides) like many instructors did.*" She further supplemented her answer by adding the addendum, "*[Sometimes] it (digital teacher) can ask questions, and [sometimes] I can ask questions. This is not a mono way but bidirectional (two-way) communication. In actual class, we all ask and answer questions, right?*"

Apart from the Q&As, some participants proposed more types of interaction and portrayed interaction should be based on situation and context. P12 suggested the use of recognizable celebrities, "*I expect more interactive content. For example, bringing historical figures or artifacts into the 3D space and letting them talk to me across several lifespans. This is such a transcendent [experience] that can enhance the classroom engagement.*" He further defined interactivity as "*the method of delivering knowledge that will be evolved with technology,*" and thus, summarized, "*content remains crucial all the time.*" Coincidentally, P10 suggested that static rules to improve interactivity are as crucial as the responsive one. He said, "*in Einstein's class (Lecture 10: VR Albert Einstein), I remember I could walk around him, but he didn't interact with me... I mean he did not even look at me, so more interaction, like turning to face me when I move, would be better.*" He also noted that this is a particular interaction in VR scenes by remarking, "*eye contact is a point [to enhance the interactive experience], so as following my path in VR as we all tend to explore and walk around in the virtual classroom.*" P7 highlighted "remember" in her definition of interaction, saying, "*if digital teachers can remember me and every single of us in the class, it would enhance the [learning] experience and reduce our anxiety about the participation.*" She further rebutted her suggestion that "*this is so hard and insecure to do so though... as we need cameras and data storing our private information to achieve this.*"

Besides, few participants with technical backgrounds were aware of the feasibility and level of difficulty in creating interactive digital teachers from a technological perspective. P15, a PhD student with more than 20 years of experience in the information technology field, expressed his concern, doubting, "*getting interaction is definitely better but as a technician, I know it's hard... It's actually impossible right?*" Another participant, P12, put himself into the developer position, suggesting the integration of large language models into the application, "*especially with advanced agents like ChatGPT [, they can] provide continuous interaction and support.*"

*4.2.2 Naturalness to Human Perception.* Naturalness was commonly found in participants' notes. Most participants reported that naturalness is the standard they apply to digital human application, consistent with our findings for RQ1. P2 repeatedly mentioned 'natural' when recalling her experience, saying, "*when I look at teachers, I consider many factors, not just teaching effectiveness, but also how natural the performance is,*" and she added that, "*naturalness is more than the term itself - I am looking for digital teacher with realistic verbal and non-verbal features, a 3D image that is acceptable and not too uncanny. Of course, the tone and speaking pace should meet teaching needs.*" P15 also noted the naturalness as a top design guidelines, and added that it should align with own cultural and educational background. Growing up in non-English speaking environment, he said, "*a too* *fluent English makes me realize that it is a fake character. No one can speak without stammer and mistake.*" He tried to quote the instances from course, saying, "*some digital teachers speak so fast, too fluent without stuttering and pause - it is way to perfect and flawless - this affects my learning, especially I am not a native speaker.*" He also defined that natural teaching contains error in oral. Several participants (P8, P12, P13, P15) tied the naturalness with the key factors defined in RQ1. P15 tried to compare the naturalness between real and digital teachers, saying, "*interestingly, human teachers also have constant fidget, but it won't affect me a lot. When it comes to digital teachers, I tend to check gesture patterns unintentionally even if some random actions are added.*"

An interesting finding lies in the definition of naturalness that goes beyond "human likeness" and "credibility". P7 tied the naturalness, human likeness, credibility, and teaching quality, saying ,"*the more natural they (digital teachers) seem, the higher the [teaching] quality, making it easier to accept the content and more trustworthy. As we encounter new teachers weekly, our quality threshold increases and we were always looking more natural teachers.*" Few participants (P7, P9, P15) stated that naturalness enhance the credibility and realism of digital teachers. They revealed that digital teachers, unlike normal digital humans, should be designed in the education context. In other words, enhancing realism and credibility should be considered as the key design guideline. P9 referred to the educational context of digital teachers, saying, "*I may prefer a more formal settings especially this is for education. I personally like formal attire in outlook and behavior.*" P7 also noted that "*if this is a linchpin course teaching core useful knowledge, I think credibility is needed to make sure students believe in him/her. The digital teachers then need to be really formal.*"

Moving from credibility, several participants stressed the importance of naturalness according to the course objectives and teachers' design styles. A common saying amongst is the necessity of exaggeration in stylized teachers and resemblance in humanoid or realistic teachers. P12 who revealed himself as the AIGC fans, emphasized that, "*different teachers for difference scenarios are better. For example, two-dimensional teacher for two-dimensional content is more appropriate.*" He further defined "*two-dimensional content*" as "*art or ACGN related study subjects such as animation.*" Furthermore, most participants (N=9) noted that "professional" and "formal" enhances the credibility as educators. For instance, P7 said, "*I trust teachers who look more like a professional real-life educators. The others seemed less realistic or too cartoonish, which felt off.*"

*4.2.3 Augmenting the Digital Teachers with more Functions.* Integrating multimedia and multi-modal instructional elements was suggested to enrich the learning experience. The majority agreed the needs in having more interactions, formats and presentations in digital teachers. P9 suggested several creative multi-modal elements. suggesting, "*I am looking for more functions, like the Liu Qiangdong's*[2] *sales. I mean referencing such formats to demonstrate in-class experiments is awesome.*" He also proposed, "*drawing and traditional whiteboard functions can allow digital teachers writing on the board as real teachers, enhancing the sense of credibility and amusement in teaching. Therefore, I expect not just videos but also teaching demonstration, white-boarding, and experiments by digital teachers!*" P14 addressed "*storytelling and gamification elements*" as the innovative multimedia approach. She further added, "*we have to be really careful on what content to be presented and make sure it is related to the class. The story of Aria is great but somehow I found it a bit off-track.*"

Besides, empowering students to engage in a broader range of tasks beyond traditional text-based learning can foster a greater sense of ownership and agency in their educational journey. P1, recalling her learning experience via YouTube and online platforms, advised that, "*personalized learning where they skip parts I already know and teach new things would be great. For example, knowing what I already know and skipping those part is the power of online learning.*" She highlighted that the customized learning path is one of the key advantages of digital teachers from her observation. P6 was excited to share that, "*digital teachers can democratize education by allowing experienced teacher to share their knowledge widely with virtual assistants. This is a great achievement for both educators and public learners.*" P15 also said, "*revisiting the digital teachers may be useful for his customized learning path.*"

[2]On 16 April, Liu Qiangdong, the founder of JD.com, made his debut in the form of a "digital human" on the JD.com. In his livestreaming, digital Liu called himself "Big brother Dong" and consumers as "brothers."

Accessibility is another key in this section. Expanding the application of asynchronous and large-scale learning modalities can further enhance the learning process. Specifically, P15 was a strong enthusiast on enhancing the digital teachers' accessibility over different platforms. He shared, "*I want the digital teacher to be approachable and appeared on different platforms.*" When asked to elaborate his answer, he added, "*The ideal part would be how the virtual teacher can assist my learning 24/7. I mean, I can text to him/her in WhatsApp; talk when I have question anytime anywhere; and even meet her in VR, Zoom, or Skype; like the J.A.R.V.I.S. in Iron Man. I believe this is the most beneficial part outweighing human instructors.*" P1 agreed that accessibility is important, explaining,"*I hope I can attend classes from home with a headset in the future, regardless of location, which would be a big improvement of having digital teachers.*"

*4.2.4 Meaningful Connection with the Study Subject.* The theme was mentioned for nine times from interviewees. Most of them agreed to have a customized digital teachers as they can enhance learning efficiency. Interestingly, interviewees revealed that these customized digital teachers should also have meaningful connection with the study subjects' topic and content. P15 commented on the division of digital and human teachers should be based on the study subject content, saying, "*Conceptual content is more suitable for digital teachers since it is scripted and more organized especially when I can revisit these structural contents, it is even better than human teachers.*" He later added that, "*for other example like the mental health or general education courses for undergraduate, it is probably fine to have anime or some cartoon agent. For these case, digital Ben [the digital version of real instructor] is not suitable since he cannot make you feel relax.*" P7 tried to put herself in teaching team's position, proposing, "*they [digital teachers] can be useful for courses that are more conceptual and less hands-on, like our social science courses. They can't fully replace real teachers but can handle some parts of the lessons. A mix of digital and real teachers would be ideal.*" P4 also associated herself to the course designing team, suggesting, "*although I think this approach may deepen some stereotypes, I think different teachers in different scenarios may be better.*" She further added the example, "*a two-dimensional teacher may give some 'two-dimensional' related explanations. For example, putting the game for anime teachers may be more appropriate than putting it here with Einstein.*" When asked about the reason, she explained that "*the Einstein we had in class is a cartoon anime character and Einstein is way too iconic. Honestly, When you have such a digital teacher, won't you feel that he is teaching a physics class?*" P10 shared his idea on VR digital teachers course design, stating, "*digital teachers is good for general courses. For humanities and arts courses, VR digital teacher can enhance the immersive experience. But for unassuming subjects like maths formula, offline classes might be better. In VR, the teacher can't see what students are doing, reducing supervision and focus.*" He further clarified that by unassuming, it was referring the repeated content that is more 'scripted' and designed ahead of schedule.

Although this was not frequently mentioned by all participants, it was observed from most of the ideal digital teachers design worksheets. Participants tended to tie digital teachers' design with the course content. It was first found from the design that there are five major types (see figure 5): (i) fictional stylized (N=2), (ii) fictional realistic/humanoid (N=4), (iii) stylized with reference (N=4), (iv) realistic/humanoid with reference/historical figure (N=3); and (v) non-human (N=2). All avatars and course design are correlated. For example, P6 designed Chongzhi Zu[3], the well-known

[3]The Chinese astronomer, inventor, mathematician, politician, and writer during the Liu Song and Southern Qi dynasties

Chinese historical figure, to teach mathematics in class, aligning the participant's own cultural background and the character's image. P4 also referenced the famous animated science fiction sitcom, *Rick and Morty*, inviting the characters to join as digital teacher teaching Multiverse concept, co-align with the course intention and outcomes. Most of the design worksheets (N=13) revealed the correlation between digital teachers' design and the course content. This highlights that the meaningful connection with the study subject is one of the crucial design guidelines.

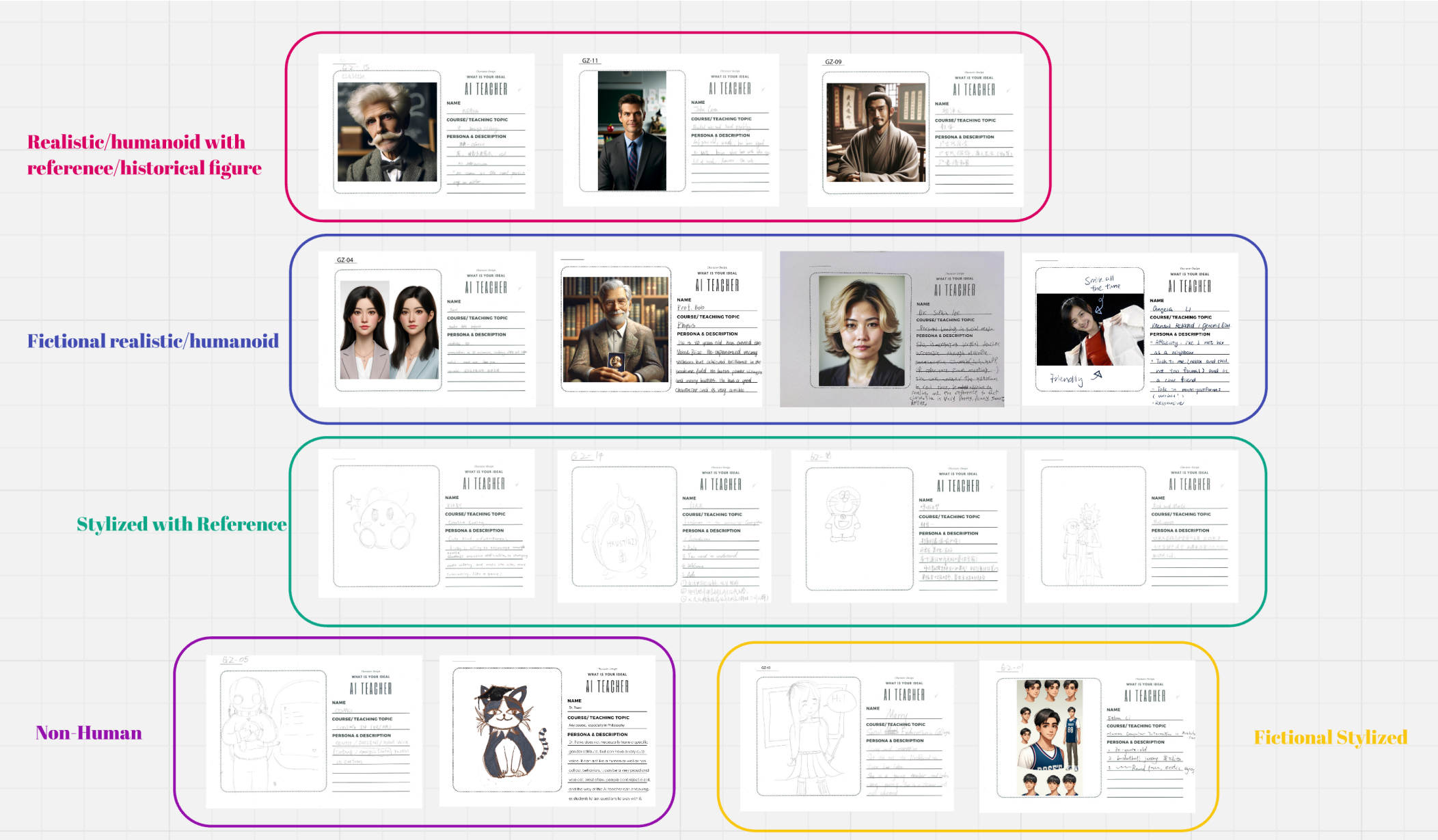

Fig. 5. The showcase of worksheets *Who is Your Ideal Digital Teacher?*. These can further be categorized into five types which most of the design are correlated with study subjects.

# 5 DISCUSSION

## 5.1 Digital Teacher as a Complementary Role

*5.1.1 Substituted.* This research do not agree either digital teachers or human instructors can replace one another. Substitutability between digital and human instructors is low since their format, practices, and classroom roles are different. Instead, AI lecturers assist professors / human teachers. Different from ordinary lecturing, where the professor explains the course content, AI lecturers can intrigue and enhance students' engagement as they can integrate the class content into one with different media for presentation. AI lecturers have different personalities and meet students in various ways, bringing an immersive learning platform to another level.

*5.1.2 Personalized.* Resemblance within their character and attitude and feel much better Regarding style, realistic or photorealistic styles can make the digital teacher seem more natural and less distracting. However, 3D stylized or comic styles, which are unreal in nature but have human-like movements, can appear awkward to students and may lead to the uncanny valley effect. Regardless of whether the style is realistic or stylized, authentic movements and expressions are crucial for maintaining students' concentration and interest. Familiar and unfamiliar playfullness of character (relationship between subjective teaching and character)

## 5.2 Enhanced Learning Experience and Engagement

*5.2.1 Media Integration.* Our research highlights the use of media integration in digital teachers. AI lecturer discusses some actual events that have happened. These can be presented in forms of videos, sound effects, historical narration, etc., in visual and auditory ways to increase students' classroom immersion. In terms of teaching experience and engagement, AI lecturers can create many unique classroom experiences from different angles. For example, the storytelling section, where students can "experience" the textbook knowledge.

*5.2.2 Diversified Choice of Contents and Practices.* At the current stage, AI lecturers cannot interact with students as we need to ensure the content accuracy. Teaching team needs to manually review the generated manuscript. Digital teachers and human lecturers have different roles: AI is not able to teach exact content with real experience. The content logical flow is convergent thinking, where they find concrete and familiar solutions to a problem. The generated content tends to favor practical over innovative solutions. They are good at elaborating prior knowledges instead of explaining and introducing new knowledges. Human instructors perform better in explaining relatively complex or new theories as they can discuss with their own personal examples and showcase more angle with real-time responses. AI teachers' uninhibited nature of the media helps students understand the content > the general teaching strategies of pictures/video + explanations by human instructors. Introducing new characters for each class keeps students excited and looking forward to lessons. Suprising factor Characters in VR teaching formats enhance students enthusiasm and engagement because of the immersive experience and gives students freedom to study on their own.

## 5.3 Accessibility and Equity

*5.3.1 Asynchronous and Personalized Education.* AI instructors promote flipped classroom and asynchronous learning. Students learn the content outside of class through pre-recorded lectures, readings, or online resources. Classroom time is then utilized for interactive activities, discussions, and problem-solving exercises. This approach allows students to engage in active learning during face-to-face sessions and seek clarification or support from instructors when needed. Promoting the accessibility, inclusivity, and flexibility behind asynchronous teaching. No longer restrict students from studying in class: students with different needs and level of understanding to certain topics can cater to own learning progress and ability. In classroom, teachers can focus more on consolidating students' understanding and applying more collaborative or active learning activity in class.

*5.3.2 Massive Learning and MOOCs.* Asynchronous is not a new method in teaching – what are the differences to have digital teachers be involved in the process? Due to the automated generation method, AI lecturers reduce the creation costs of the teaching team or production team. A computer can generate different AI lecturers without even using any shooting and lighting tools. From an educator's perspective, AI lecturers have the potential to expand the scale of education delivery by reaching a large number of students at the same time. This scalability can also help solving the problem of teacher shortages, especially in some undeveloped areas.

## 5.4 Teachers For All: Design Considerations

90% of our interviewees (students in the class, N=20) want to have (and looking forward to) digital teachers in the class and prefer engaging with DHs in more activities. When the AI lecturer is busy discussing the content, will the teacher be too "idle"? Human teachers should make use of the class and provide corresponding assistance. More engagement should be involved: VR storytelling by digital teachers and in-class collaborative learning should be designed. Teachers can adjust teaching methods and incorporate new information and research results to ensure students receive the latest relevant educational content and use more examples to illustrate more details.

### 5.5 Limitations and Future Work

This study, while providing valuable insights into the impact of digital teachers on students' learning experiences, is not without its limitations. Firstly, the data collection method relied heavily on interviews, which may introduce biases related to self-reporting and the subjective nature of participant responses. Future research could benefit from incorporating a mixed-methods approach, including quantitative data such as performance metrics and engagement analytics, to provide a more comprehensive understanding.

Secondly, the sample size for this study was relatively small, limiting the generalizability of the findings. A larger and more diverse sample would allow for more robust conclusions and a better understanding of how different demographic groups interact with digital teachers.

Additionally, the current design of the digital teacher is not fully utilized and lacks comprehensiveness. The digital teachers in this study were neither interactive nor responsive, which significantly limits their effectiveness. Enhancements in these areas are crucial for realizing the full potential of digital teachers in educational settings.

Future research should focus on leveraging the design of digital teachers to create more personalized, interactive, and responsive educational tools. One promising direction is the development of personalized avatars that can adapt to individual learning styles and preferences. These personalized digital teachers could provide tailored feedback and support, thereby enhancing the learning experience.

Interactivity is another critical area for future work. Developing digital teachers that can engage in bidirectional interactions with students will make learning more engaging and effective. This could include real-time question-and-answer sessions, adaptive learning paths, and interactive simulations.

Enhancing accessibility is also paramount. Future iterations of digital teachers should incorporate multiple communication channels, including social media platforms, to provide students with more ways to interact and seek assistance. This multi-channel approach can make digital teachers more accessible and user-friendly.

Moreover, integrating digital teachers into larger platforms such as Massive Open Online Courses (MOOCs) could significantly expand their reach and impact. By doing so, digital teachers can cater to a broader audience, providing high-quality education to learners worldwide. This scalability can help address educational inequities and ensure that more students have access to the resources they need to succeed.

In conclusion, while this study has highlighted important aspects of digital teachers, there is considerable scope for improvement and expansion. By addressing the current limitations and focusing on future enhancements, we can move closer to realizing the vision of "Teachers for All," ensuring a more inclusive, engaging, and effective educational experience for all learners.

## 6 CONCLUSION

"Teachers for All" envisions a future where technology-enhanced education addresses the pressing issues of a lack of trained teachers and inadequate educational materials, enhancing the personalized learning engagement. This paper explores how variations in key factors of digital teachers impact students' experiences and identified potential areas for improvement.

Our findings indicate that verbal and non-verbal cues play a significant role in shaping students' learning experiences. While verbal expressions, such as clear and natural-sounding voices, are preferred, non-verbal cues like facial expressions and body language are crucial for maintaining engagement and credibility. Characters that resemble real humans or familiar figures, such as a digital version of a professor or Einstein, enhance trustworthiness and engagement. However, the uncanny valley effect can occur with unrealistic or overly stylized characters, underscoring the importance of authenticity and realism. Furthermore, resemblance within teachers and their attitude is recommended. Due to the relationship between subjective teaching and character design, the style of digital teachers should be consistent with the teaching subjects. Particularly, stylized or anime characters, implying the higher playfulness, should be assigned to general education or courses that are more related to mental health maintenance; whilst realistic or hyper-realistic teachers should be should be designed in the education context, of which enhancing the realism and credibility is the key design guideline.

The introduction of new characters and storytelling methods can enhance student enthusiasm and engagement as surprising factor. The use of VR teaching formats, particularly with well-known figures, has shown to be highly effective in creating an immersive learning experience. However, the repetitive nature of some gestures and the inability to easily revisit VR classes highlight areas for improvement.

In terms of future directions, our research suggests that interactivity and personalization are paramount. Students expressed a strong need for bidirectional interaction, allowing for a more engaging and responsive learning environment. The potential for personalized digital teachers tailored to individual learning styles could significantly enhance educational outcomes.

Moreover, digital teachers have the potential to complement traditional teaching by taking over certain educational duties, thus reducing the teaching load on human educators. The integration of digital teachers with VR and MR technologies can provide a more immersive and interactive learning environment, enhancing the overall educational experience.

Accessibility and equity are also critical considerations. Digital teachers can help democratize high-quality education, making it accessible to a broader audience. The use of multiple and cross-social media platforms can further enhance the accessibility and effectiveness of digital teachers.

In conclusion, the future of education lies in the effective integration of AI and Metaverse technologies. By addressing current limitations and embracing innovative practices, we can move closer to the vision of "Teachers for All." This will ensure that every learner has access to the resources and support they need, ultimately fostering a more inclusive, engaging, and personalized educational landscape.

## 7 ACKNOWLEDGEMENT

We would like to express our sincere gratitude to the Guangzhou Municipal Nansha District Science and Technology Bureau for their support under Contract No. 2022ZD012. Their commitment to fostering innovation in education has been instrumental in the advancement of our research on digital teachers.

Additionally, we acknowledge the generous sponsorship provided by the HSBC project L0562, which has significantly contributed to our study. This funding has enabled us to explore the integration of digital teaching methodologies, enhancing our understanding of their impact on higher education.

We extend our heartfelt thanks to the students who participated in our research. Their willingness to engage with the digital teachers and provide meaningful feedback has been invaluable in shaping our findings and insights. Their contributions have enriched our study and underscored the importance of student perspectives in the evaluation of educational technologies.

We are grateful for the collaborative efforts and resources provided by both organizations, which have been vital to the success of our research.

# A APPENDIX A: INTERVIEW MATERIALS

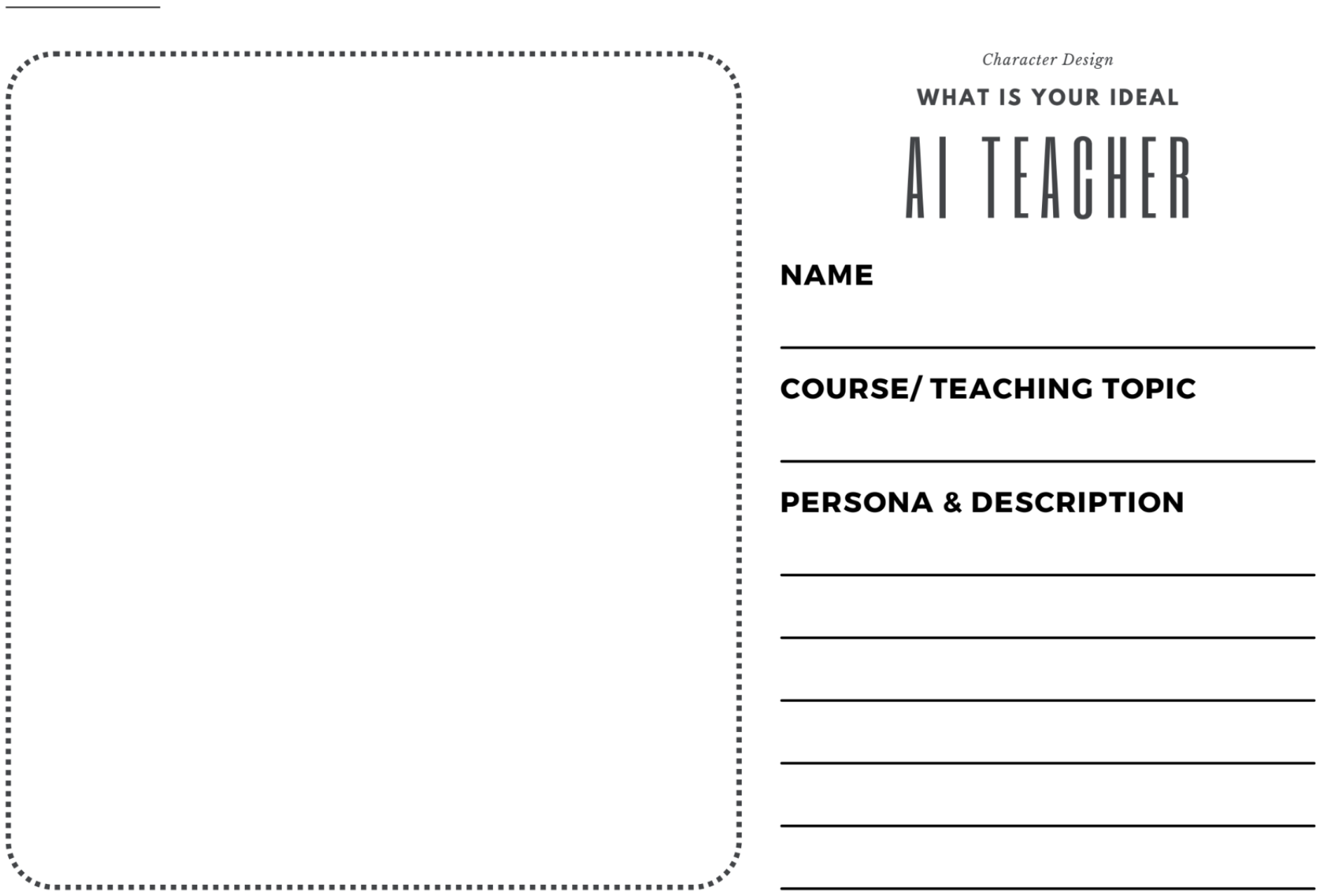

Fig. 6. The interview worksheets *Who is Your Ideal Digital Teacher?*.

## B APPENDIX B: RQ1

| Themes | Frequency | Keynotes and Highlights |
|---|---|---|
| Voice and Verbal Expression | 22 | Students mentioned several elements of the voice and verbal expressions of digital teachers, including pitch, tone, accent, fluency, and the speed of speaking. Overall, pitch and tone do matter for digital teachers' natural delivery. Students think that digital teachers can have an accent and do not need to be very fluent in speaking, as real teachers do too, making the experience feel like a real-life lecture. Speaking speed does not matter for most students, but for those who find it important, digital teachers should not speak too quickly and should include pauses to make it easier to follow content, especially for non-native English speakers. |
| Non-verbal Cues | 28 | Most students believe that non-verbal cues such as body movements, postures, and facial expressions impact their learning experience. Rigid, stiff movements make students uncomfortable and distract from the lecture, while smooth movements help focus on the content. Some students prefer full-body digital teachers for interactive movements, while others find it distracting. Repetitive gestures without variation can make students feel bored. Clear, non-repetitive facial expressions leave a stronger impression on students. |
| Appearance and Character Choices | 57 | Students' learning experiences are influenced by the style, familiarity, and attributes like gender, age, and ethnicity of digital teachers. Realistic styles with natural movements resembling real people are preferred. Photo-realistic 2D styles can also be impressive. Familiar characters increase comfort and engagement with digital teachers. |
| Variety and Novelty | 30 | The format of digital teacher presentations, such as VR environments, offers more movement freedom and immersion, creating a refreshing experience. Edited videos with illustrative images and dynamic media are rated as effective in conveying information. Rotating different digital teachers for each class increases anticipation and boosts learning motivation. |

## C APPENDIX B: RQ2

| Themes | Frequency | Keynotes and Highlights |
|---|---|---|
| Situation-Aware Interactivity | 22 | Given the existing unresponsive versions, students expressed a strong desire for digital teachers to exhibit a deeper understanding of the learning context and to adapt their responses accordingly. Participants highlighted the importance of digital teachers being able to reply according to the needs of students, particularly the tailored interactivity in Q&As. Students preferred to allow digital teachers to interrupt and ask questions rather than simply answer. |
| Naturalness to Human Perception | 19 | While impressed by the technological capabilities of digital teachers, students noted the need for these digital instructors to be more natural and relatable. Participants suggested that enhancements to the body language, facial expressions, eye contact, and tone of voice are necessary to make digital teachers more lifelike and less repetitive. Some also suggested the needs of "flaws" in conversation. Naturalness goes beyond "human likeness" is emphasized. "Resemble" and "exaggerated" were the words they mentioned most frequently when discussing their expectations for realistic and stylized/anime teachers respectively. Naturalness is further linked to the realism and credibility of digital teachers. Participants emphasized that digital teachers need to be in a formal outlook to fit in the scholarly image and academic setting. |
| Augmenting the Digital Teacher with More Functions | 19 | Students expressed a desire for digital teachers to offer a more comprehensive suite of features and support, extending beyond the core instructional role to provide creative teaching methods, such as storytelling and experiment demonstrations. Participants highlighted the potential for digital teachers to provide personalized study plans, facilitate peer-to-peer collaboration, offer real-time feedback and guidance, and even serve as mentors for academic and mental development. Students revealed that they want to take ownership of the learning process with the assistance of digital teachers, working on active learning tasks that go beyond text and traditional pedagogical approaches. Exploring the integration of media and interaction techniques was suggested to further bridge the gap between digital teachers and human-to-human interactions. |
| Meaningful Connection with the Study Subject | 9 | While appreciating the convenience and accessibility of digital teachers, some students expressed a need for these virtual instructors to foster a stronger sense of connection between the learner and the subject matter. Participants suggested that digital teachers should be designed considering the course content and CILOs. Participants emphasized the needs of dividing works for human and digital teachers. In particular, digital teachers were seen as potentially taking over certain teaching duties for compulsory and general education, thereby leaving more complex content for human instructors. |